\documentclass[12pt,a4paper]{article}
\usepackage[cp1251]{inputenc} 
\usepackage[english]{babel}
\usepackage[final]{graphicx}
\usepackage[colorlinks=false]{hyperref}

\begin{document}

\title{Nuclear spin-orbit interaction and T-odd angular correlations in ternary fission}
\author{A. L. Barabanov\\
{\normalsize\it NRC ''Kurchatov Institute'', Moscow 123182, Russia; Moscow Institute of}\\
{\normalsize\it Physics and Technology, Dolgoprudny 141700, Moscow Region, Russia}}
\date{}
\maketitle

\begin{abstract}
T-odd angular correlations in ternary fission of $^{233}$U and $^{235}$U nuclei by slow polarized neutrons are not related to TRI (time reversal invariance) violation, but are caused by an effective spin-orbit interaction in the final state.
\end{abstract}

\section{Introduction}
In the decay of a polarized particle N with a spin $s_N$ to three or more particles with momenta ${\bf p}_1$, ${\bf p}_2$,\ldots, T-odd angular correlation $({\bf s}_N [{\bf p}_i\times {\bf p}_j])$ may arise. The correlation appears in the first order on the interaction which is responsible for the decay provided this interaction violates time reversal invariance (TRI) (see, e.g., \cite{1}). So, in particular, measurements of the amplitude D of the correlation $({\bf s}_n [{\bf p}_e\times {\bf p}_{\tilde\nu}])$ in free neutron beta-decay, n\,$\to$\,p\,+\,e\,+\,$\tilde\nu$, are used for years to search TRI-violating weak forces (see \cite{2} for the best today's limit on D in the free neutron decay).

In the second order the ''usual'' forces (so-called final-state interactions), which have nothing to do with TRI violation, can generate T-odd angular correlations \cite{1}. Therefore, in general, the detection of T-odd correlation $({\bf s}_N [{\bf p}_i\times {\bf p}_j])$ in the decay of polarized particle N does not point out the TRI violation. Notice that the expected contributions of final-state interactions in free neutron decay are small in comparison with the current accuracy of measurement of the factor D \cite{2} (''false'' TRI-violating effects can also be separated from ''true'' ones owing to different dependence on energy of decay products).

\begin{figure}[h]
\label{f1}
\begin{center}
\mbox{\includegraphics*[scale=0.9]{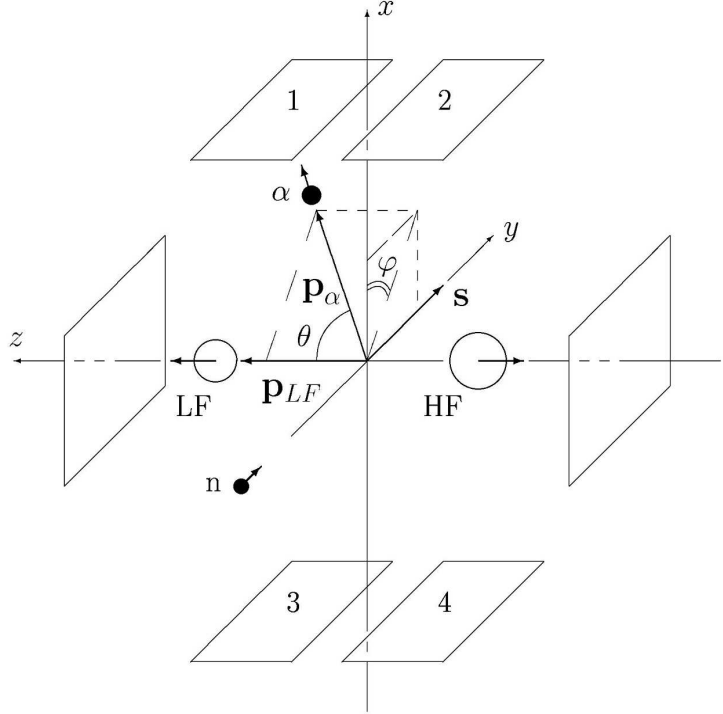}}
\caption{Scheme of detection of T-odd correlations in ternary fission: n -- neutron with spin ${\bf s}$, $\alpha$ -- $\alpha$ particle with momentum ${\bf p}_{\alpha}$, LF -- light fragment with momentum ${\bf p}_{LF}$, HF~-- heavy fragment; 1, 2, 3 and 4 -- $\alpha$-particle detectors.}
\end{center}
\end{figure}

In the first publications \cite{3,4} devoted to detection of T-odd angular correlation $({\bf s} [{\bf p}_{LF}\times {\bf p}_{\alpha}])$ in ternary fission of nuclei $^{233}$U by slow (s-wave) polarized neutrons, possible relation to TRI violation has been allowed, therefore the above correlation was labelled as TRI effect. Here ${\bf s}$, ${\bf p}_{LF}$, ${\bf p}_{\alpha}$ are the spin of an incident neutron, the momenta of light fragment and $\alpha$ particle (see Fig.~\ref{f1}). The $\alpha$ particle is emitted practically simultaneously with fragments at ternary nuclear fission. The compound nucleus $^{234}$U$^*$ is the decaying polarized particle N, and its spin ${\bf s}_N\equiv {\bf J}$ is directed along the spin ${\bf s}$ of the polarized neutron captured by the target nucleus $^{233}$U.

However, nowadays there is no doubt that T-odd correlations in ternary fission are caused by strong final-state interactions, therefore, they are not related to TRI violation. Unfortunately, the confusing label ''TRI effect'' still remains in use. In what follows, we name the asymmetry that is related to the correlation $({\bf s} [{\bf p}_{LF}\times {\bf p}_{\alpha}])\equiv ({\bf p}_{\alpha} [{\bf s}\times {\bf p}_{LF}])$ as T-effect (transverse asymmetry, i.e., asymmetry of $\alpha$-particle emission with respect to the plane formed by the vectors ${\bf s}$ and ${\bf p}_{LF}$).
\vspace{0.5cm}

\section{Model with spin-orbit interaction}

Right after the correlation $({\bf s} [{\bf p}_{LF}\times {\bf p}_{\alpha}])$ was found, I have advanced a schematic explanation of the mechanism of its formation. It was shown that the asymmetry of $\alpha$-particle emission can be caused by an interaction of spin ${\bf J}_0$ of the system of two fragments separating from each other and the orbital momentum ${\bf l}$ of already separated $\alpha$ particle, i.e., spin-orbit interaction $\sim {\bf l}{\bf J}_0$. I discussed this hypothesis with the authors of the works \cite{4,5}, that is why the model of spin-orbit interaction was mentioned in \cite{4,5}. The most complete description of the model was presented at the conference in Dubna in 2001 and published in its proceedings \cite{6}.

A similar model based on the interaction $\sim {\bf l}{\bf J}_0$, named, however, ''Coriolis interaction'', was proposed by V.E.~Bunakov and S.G.~Kadmensky~\cite{7}. They first noted that the scheme \cite{6} ''makes it possible, in principle, to cal\-cu\-late the asymmetry coefficient in ternary fission for alpha particles\ldots '', but later rejected it on the mistaken assumption (groundlessly attributed to the author of \cite{6}) ''that the spin ${\bf J}_0$ can be replaced by the total spin of the nucleus undergoing fission, ${\bf J}$\ldots '' Actually, I have assumed in \cite{6} that the initial orbital momentum ${\bf l}$ of $\alpha$ particle is small (by the way, this smallness has been widely used in \cite{7}), therefore, ${\bf J}={\bf J}_0+{\bf l}$ and ${\bf J}_0$ are approximately equal. Thus, the arguments advanced in \cite{7} against the spin-orbit model are completely wrong.

Furthermore, the Coriolis interaction exists only in a noninertial system of coordinates. Meanwhile, the asymmetry of $\alpha$-particle emission is observed in the inertial laboratory system and, hence, it is caused by the interaction that takes place in the inertial system. On the other hand, if in \cite{6} and \cite{7} the interaction $\sim {\bf l}{\bf J}_0$ is the same, so will be the conclusions. In particular, it was stated in \cite{6}: ''No strong dependence of the asymmetry on the angle of $\alpha$-particle emission with respect to the fission axis is predicted by the model in accordance with the experimental data''. Then we find in \cite{7}: ''The T-odd-asymmetry coefficient\ldots will then be weakly dependent on the angle $\theta_r$, and this is consistent with its experimental properties''. Similarly, a crucial role of mixing of the states of $\alpha$ particles with even and odd orbital momenta resulting from the charge asymmetry between light and heavy fragments was established in \cite{6}. In turn, the authors of \cite{7} come to the conclusion: ''The proof of the fact that T-odd correlations in the ternary fission of nuclei arise only in fission modes that are asymmetric in fragment charges and masses is an important result of the present study''.

In \cite{6}, it was shown  that the angular correlation $({\bf p}_{\alpha} [{\bf s}\times {\bf p}_{LF}])$ is not enough for description of T-effect. Indeed, let us direct an axis $y$ along the vector~${\bf s}$, an axis $z$ along the vector ${\bf p}_{LF}$, so the vector $[{\bf s}\times {\bf p}_{LF}]$ is directed along an axis $x$ (see Fig.~\ref{f1}). The calculation \cite{6} gave the following expression for the angular and energy distributions of $\alpha$~particles:
\begin{equation}\label{1}
\frac{dw}{d\Omega dE_{\alpha}}=W_0(\theta)+
p(J)\cos\varphi\left(W_1(\theta)+W_2(\theta)\right),
\end{equation}
where
\begin{equation}\label{2}
W_0(\theta)=\sum\limits_{Q=0,1,2,\ldots}(2Q+1)a_Q(E_{\alpha})P_Q(\cos\theta),
\end{equation}
\begin{equation}\label{3}
W_1(\theta)=\sum\limits_{Q=1,3,\ldots}(2Q+1)b_Q(E_{\alpha})P_Q^1(\theta),
\end{equation}
\begin{equation}\label{4}
W_2(\theta)=\sum\limits_{Q=2,4,\ldots}(2Q+1)b_Q(E_{\alpha})P_Q^1(\theta).
\end{equation}
The functions $P_Q(\cos\theta)$ and $P_Q^1(\theta)=\sin\theta\left(dP_Q(\cos\theta)/d\cos\theta\right)$ are usual and associated Legandre polynomials, $p(J)$ is the polarization of compound nucleus with spin $J$. Explicit expressions for the factors $a_Q(E_{\alpha})$ and $b_Q(E_{\alpha})$ are presented in \cite{6}. In the second term at the right-hand side of Eq.(\ref{1}) the contributions from odd and even $Q$ are separated ($W_1$ and $W_2$, respectively). This second term results from the spin-orbit interaction and, being proportional to $\cos\varphi$ (the angle $\varphi$ on the plane $(x,y)$ is reckoned from the axis $x$ -- see Fig.~\ref{f1}) describes the asymmetry of $\alpha$-particle emission along ($\varphi=0$) and contrary ($\varphi=\pi$) to the axis $x$, i.e., T-effect.

It is easy to see that only the first contribution to this term, corresponding to $Q=1$, gives the correlation $\cos\varphi\,\sin\theta\equiv ({\bf n}_{\alpha} [{\bf n}_s\times {\bf n}_{LF}])$, where ${\bf n}_{\alpha}$, ${\bf n}_s$, ${\bf n}_{LF}$ are unit vectors along ${\bf p}_{\alpha}$, ${\bf s}$, ${\bf p}_{LF}$. However, in general, the contributions from $Q=2,3,\ldots$ to the functions $W_1$ and $W_2$ are important in the same way as similar contributions to the function $W_0$. Indeed, the function $W_0$ describes a rather narrow angular distribution of $\alpha$ particles with respect to the fission axis just due to account of a great number of contributions.

In real experiments, $\alpha$ particles are registered by detectors placed in the plane $(x,z)$: detectors 1 and 2 are situated at the angles $\theta$ and $\pi-\theta$ to the axis $z$ in the positive side of the axis $x$, i.e., $\varphi= 0$, and detectors 3 and 4 are situated at the same angles $\theta$ and $\pi-\theta$ to the axis $z$ in the negative side of the axis $x$, i.e., $\varphi=\pi$ (see Fig.~\ref{f1}). Thus, two different partial asymmetries of $\alpha$-particle emission can be introduced: for the angle $\theta$ ($D_{13}$ according to the detectors 1 and 3) and for the angle $\pi-\theta$ ($D_{24}$ according to the detectors 2 and 4). When the incident neutrons are not polarized, $p(J)=0$ and $D_{13} = D_{24} = 0$. However, in this case the detectors 1 and 3 register more $\alpha$ particles than the detectors~2 and 4 owing to Coulomb repulsion that is stronger from the heavy fragment than from the light one, that is why the maximum of $\alpha$-particle angular distribution is shifted to the angle $\theta\simeq 83^0$.

In the measurements \cite{4,5} it was found that $D_{13}\simeq D_{24}$ for target nuclei $^{233}$U. This~fact caused the sceptical relation to the spin-orbit model. The point is that simplified semiclassical picture corresponding to this model (see details in \cite{6}) allows the following interpretation: ''\ldots in case the spin-orbit interaction is at work, the correlation coefficient $D$ should have opposite signs for angles smaller or larger than the average angle 83$^0$ \ldots ''~\cite{5}. Thus, according to the authors of \cite{5}, in the model with spin-orbit interaction one should expect $D_{13}\simeq -D_{24}$.

This scepticism, however, is absolutely unreasonable. Using the realistic quantum model instead of naive semiclassical picture, we obtain Eqs.(\ref{1})-(\ref{4}), where
\begin{equation}\label{5}
W_1(\theta)=W_1(\pi-\theta),\quad
W_2(\theta)=-W_2(\pi-\theta).
\end{equation}
Hence, according to the data \cite{4,5} (for the target nuclei $^{233}$U) the function $W_1$ dominates, while the function $W_2$ is suppressed for some reasons. In other words, T-effect, when \mbox{$D_{13}\simeq D_{24}$}, corresponds to dominance of $W_1$ and suppression of $W_2$. However, for some other target nuclei one may expect the opposite situation: $W_1$ is suppressed, $W_2$~dominates, then $D_{13}\simeq -D_{24}$ in casual accordance with the semiclassical picture.

The surprising thing is that this was found in the latter half of 2000 in ternary fission of nuclei $^{235}$U by slow neutrons \cite{8,9}\,! Unfortunately, the scepticism with respect to the spin-orbit model has not been revised. The authors of \cite{8,9} related the new type of asymmetry with nuclear ''rotation'' before scission and labelled it as ''ROT effect''. Shorter label \mbox{''R-effect''} seems more preferable and will be used in what follows. Obviously, R-effect, when $D_{13}\simeq -D_{24}$, corresponds to suppression of $W_1$ and dominance of $W_2$.

It should be emphasized that Eqs.(\ref{1})--(\ref{4}) describe both T-effect and R-effect, but they has been obtained in \cite{6} before the measurements \cite{8,9} were performed. In the general case, the contributions from $W_1$ and $W_2$ can be of the same order (both T-effect and R-effect will be observed).

R-effect corresponds to T-odd angular correlation $({\bf p}_{\alpha} [{\bf s} \times {\bf p}_{LF}]) ({\bf p}_{\alpha} {\bf p}_{LF})$. However, in general, this correlation is not enough for description of R-effect. Indeed, let us take the first contribution to $W_2(\theta) $ that corresponds to $Q=2$. It has the form $\cos\varphi\,\sin\theta\,\cos\theta\equiv ({\bf n}_{\alpha} [{\bf n}_s\times {\bf n}_{LF}])({\bf n}_{\alpha} {\bf n}_{LF})$. But the next contributions ($Q=4,6\ldots $) to $W_2$ may also be significant. 

\section{Effective spin-orbit interaction}

Spin ${\bf J}_0$ of two fragments separating from each other transforms to the sum ${\bf J}_{LF}+{\bf J}_{HF}+{\bf L}$ after scission, where ${\bf J}_{LF}$ and ${\bf J}_{HF}$ are spins of light and heavy fragments, and ${\bf L}$ is a relative orbital momentum. So, the question arises: what are the relative contributions to the spin-orbit interaction ${\bf l}{\bf J}_0$ from different parts of ${\bf J}_0$ after scission?

Target spin effects in elastic scattering of $^3$He on the adjacent nuclei $^{60}$Ni ($I=0$) and $^{59}$Co ($I=7/2$) were studied in \cite{10}. Significant difference in angular distributions in backward scattering was related to the interaction $\sim {\bf l}{\bf I}$. However, later \cite{11} there was proposed another interpretation of these data. It was assumed that the ground state of odd nucleus, i.e., $^{59}$Co, may be presented ''as a single hole in an orbit $j$ coupled to states of the core with spin $L$, where the core is taken to be the adjacent even nucleus'', i.e.,~$^{60}$Ni. Thus, the wave function of the ground state of $^{59}$Co was expressed via the wave functions of ground and excited states of $^{60}$Ni (see Eq.(1) in \cite{11}),
\begin{equation}\label{6}
|{\rm odd},IM\rangle=\alpha|({\rm even},0),j=I;IM\rangle+
\sum\limits_{L,j}\beta_{Lj}|({\rm even},L),j;IM\rangle,
\end{equation}
while for the elastic cross sections it was obtained (see Eq.(2) in \cite{11}):
\begin{equation}\label{7}
\sigma_{{\rm el}}({\rm odd})\simeq
\sigma_{{\rm el}}({\rm even})+\sum\limits_L\frac{(2\alpha\beta_{LI})^2}{2L+1}\,\sigma_{{\rm inel}}({\rm even},0^+\to L^+).
\end{equation}
Therefore, an enhancement of elastic scattering on the odd nucleus that has been found in~\cite{10} and interpreted there as manifestation of spin-orbit interaction $\sim {\bf l}{\bf I}$ can be explained by a scattering with virtual rotational excitation (in particular, Coulomb excitation) of the even nucleus inside the odd one. Obviously, such a mechanism can be called the effective interaction of the target spin ${\bf I}={\bf L}+{\bf j}$ and the relative orbital momentum~${\bf l}$ of the target and projectile particles.

In ternary fission, the Coulomb repulsion between $\alpha$ particle, on the one hand, and fragments, on the other hand, strongly influences the orbital momenta ${\bf l}$ (of $\alpha$ particle with respect to the center of mass of two fragments) and ${\bf L}$ (of two fragments) and, thus, may manifest itself as an effective $\sim {\bf l}{\bf L}$ interaction. This is a possible reply to the question posed at the beginning of this section.

This hypothesis is supported by the general expression for differential cross section of ternary fission which can be obtained in the framework of three-body approach with the use of hyperspherical harmonics. Let $\mbox{\boldmath $\rho$}_{LF}$ be the position of the light fragment with respect to the heavy one, and $\mbox{\boldmath $\rho$}_{\alpha}$ be the position of the $\alpha$ particle with respect to the center of mass of two fragments. The lengths of these vectors may be presented in the form $\rho_{LF}=\rho\sin\vartheta$ and $\rho_{\alpha}=\rho\cos\vartheta$, where $\rho$ is the hyperradius and $\vartheta$ is the hyperangle. The three-body final state of two fragments and $\alpha$ particle with definite quantum numbers, $L$, $l$, total spin $F$ of two fragments, total spin $J_0$ and hypermoment $N=2n+L+l$ ($n=0,1,2,\ldots$), may be described by the wave function that is proportional to hyperspherical harmonics.

We consider here ternary fission of a target nucleus with spin $I$ induced by s-wave neutron with de Broglie wave-length $\lambda$. Differential cross section is of the form 
\begin{equation}\label{8}
\begin{array}{l}
{\displaystyle\frac{d\sigma_f}{d\Omega_{LF}\,d\Omega_{\alpha}}}=
{\displaystyle\frac{\pi(\lambda/2\pi)^2}{(4\pi)^2}}
{\displaystyle\sum_{JJ'}}\,g_{J'}
{\displaystyle\sum_{Q=0,1}}U(IsJ'Q,Js)\,
\tau'_Q(s)\times{}\\
{\displaystyle\sum_{LL'll'nn'J_0J'_0F}}
S_J(0\frac{1}{2}\to LlNJ_0F)S^*_{J'}(0\frac{1}{2}\to L'l'N'J'_0F)
{\displaystyle\sum_{\Lambda_f\Lambda_{\alpha}}}
\hat\Lambda_f \hat\Lambda_{\alpha} \hat{l'}\hat{J'_0}\hat{J_0}
\times{}\\
C^{L0}_{L'0\Lambda_f0}
C^{l0}_{l'0\Lambda_{\alpha}0}
U(FJ'_0L\Lambda_f,L'J_0)
\left\{\begin{array}{ccc}
J & J_0 & l\\
J' & J'_0 & l'\\
Q & \Lambda_f & \Lambda_{\alpha}
\end{array}\right\}
\phi^Q_{\Lambda_f\Lambda_{\alpha}}({\bf n}_s,{\bf n}_{LF},{\bf n}_{\alpha})
\times{}\\[\bigskipamount]
\left(\int\limits_0^{\pi/2}
(\sin\vartheta)^{L+L'+2}
(\cos\vartheta)^{l+l'+2}
P_n^{L+\frac{1}{2},\,l+\frac{1}{2}}(\cos 2\vartheta)
P_{n'}^{L'+\frac{1}{2},\,l'+\frac{1}{2}}(\cos 2\vartheta)\,d\vartheta\right),
\end{array}
\end{equation}
where $g_J=(2J+1)/((2s+1)(2I+1))$, $\tau'_Q(s)$ are orientation spin-tensors for the incident neutrons ($\tau'_0(s)=1$, $\tau'_1(s)=p/\sqrt{3}$, $p$ is the neutron polarization), $U$ are normalized Racah functions, $S$ are elements of S-matrix (transition amplitudes), $\hat{a}=\sqrt{2a+1}$, $P_n^{\alpha\beta}$~are Jacobi polynomials, and
\begin{equation}\label{9}
\phi^Q_{\Lambda_f\Lambda_{\alpha}}({\bf n}_s,{\bf n}_{LF},{\bf n}_{\alpha})=
(4\pi)^{3/2}{\displaystyle\sum_{q\lambda_f\lambda_{\alpha}}}
C^{Qq}_{\Lambda_f\lambda_f\Lambda_{\alpha}\lambda_{\alpha}}
Y^*_{Qq}({\bf s})
Y_{\Lambda_f\lambda_f}(\mbox{\boldmath $\rho$}_{LF})
Y_{\Lambda_{\alpha}\lambda_{\alpha}}(\mbox{\boldmath $\rho$}_{\alpha}),
\end{equation}
are invariant spherical functions (see \cite{12}). The key T-odd angular correlations (T- and R-effects) arise as
\begin{equation}\label{10}
\phi^1_{11}({\bf n}_s,{\bf n}_{LF},{\bf n}_{\alpha})\!\sim\!
({\bf n}_{\alpha}[{\bf n}_s\times {\bf n}_{LF}]),\quad\!\!
\phi^1_{22}({\bf n}_s,{\bf n}_{LF},{\bf n}_{\alpha})\!\sim\!
({\bf n}_{\alpha}[{\bf n}_s\times {\bf n}_{LF}])({\bf n}_{\alpha}{\bf n}_{LF}).
\end{equation}

\section{Conclusion}

T- and R-effects in ternary fission are caused, apparently, by an effective spin-orbit interaction of the type $\sim {\bf l}{\bf L}$ owing to strong Coulomb repulsion between $\alpha$ particle and fragments in the final state.
\bigskip

This work was supported in part by grant NS-215.2012.2 from MES of Russia.

\end{document}